\documentclass[aps,prl,reprint,superscriptaddress,twocolumn]{revtex4-2}
\usepackage[utf8]{inputenc}
\usepackage{amsmath}
\usepackage{graphicx}
\usepackage{lipsum}
\usepackage{physics}  
\usepackage{placeins}
\usepackage{verbatim}
\usepackage{CJKutf8}
\usepackage{wasysym}
\usepackage{xcolor}
\usepackage[normalem]{ulem}
\usepackage{hyperref}
\usepackage{refcount}
\usepackage{microtype}

\newcommand{\bs}{\boldsymbol}
\newcommand{\lsi}{LSI, CNRS, CEA/DRF/IRAMIS, \'Ecole Polytechnique, Institut Polytechnique de Paris, F-91120 Palaiseau, France}
\newcommand{\etsf}{European Theoretical Spectroscopy Facility (ETSF)}
\newcommand{\ccq}{Center for Computational Quantum Physics, Flatiron Institute, 162 5th Avenue, New York, New York, USA}
\newcommand{\hofstra}{Department of Physics and Astronomy, Hofstra University, Hempstead, New York 11549, USA}
\newcommand{\soleil}{Synchrotron SOLEIL, L'Orme des Merisiers, Saint-Aubin, BP 48, F-91192 Gif-sur-Yvette, France}

\begin{document}

\title{Signature of Short-Range Order in Static Response of the Three-Dimensional Electron Gas}

\author{Muhammed H. Güneş}
\affiliation{\lsi}%
\affiliation{\etsf}%

\author{Yubo Yang \begin{CJK}{UTF8}{gbsn} (杨煜波)\end{CJK}}
\affiliation{\hofstra}%
\affiliation{\ccq}%

\author{Vitaly Gorelov}
\affiliation{\lsi}%
\affiliation{\etsf}%

\author{Miguel A. Morales}
\affiliation{\ccq}%

\author{Matteo Gatti}
\email{matteo.gatti@polytechnique.fr}%
\affiliation{\lsi}%
\affiliation{\etsf}%
\affiliation{\soleil}%

\author{Lucia Reining}
\email{lucia.reining@polytechnique.fr}%
\affiliation{\lsi}%
\affiliation{\etsf}%

\author{Shiwei Zhang}
\email{szhang@flatironinstitute.org}%
\affiliation{\ccq}%

\date{\today}

\begin{abstract}
    The three-dimensional electron gas is a fundamental model in condensed matter physics and quantum chemistry, and the exchange-correlation energy derived from it serves as the starting point of \textit{ab initio} computations of materials. However, the wave-vector-dependent response, and hence the static local field factor $G(q)$, has remained without accurate ground-state benchmark in the strongly coupled regime for three decades. Using diffusion Monte Carlo, we calculate the static density-density response function and $G(q)$ across the liquid phase and find a pronounced structure in $G(q)$ at intermediate wave vectors, already visible at metallic densities and growing with increasing interaction strength. We identify it as a fingerprint of short-range order by showing that it is required to reproduce the static structure factor. Our parametrization, valid in the entire liquid phase, predicts a low-energy resonant mode inside the particle-hole continuum.
\end{abstract}

\maketitle
The ground-state energy of the homogeneous electron gas, computed to high accuracy by quantum Monte Carlo (QMC)~\cite{CEPERLEY1980,TANATAR1989,FOULKES2001}, is the input from which approximations to density functional theory (DFT) are built~\cite{KOHN1965,Perdew1996,Sun2015,Alonso1978,Gunnarsson1979,Cuevas-Saavedra2012,Aouina2025}, and with them much of modern \textit{ab initio} materials science. How the same system responds to a perturbation is known far less precisely. Much of this information is carried by the static linear density response function $\chi(q)$ and the closely related local field factor $G(q)$, both of which have attracted sustained theoretical interest~\cite{KUGLER1975,SINGWITOSI1981, GIULIANI2005,PINES1966}. The latter encodes the full wave-vector-resolved exchange and correlation effects and enters directly into descriptions of dielectric screening, effective electron--electron interactions~\cite{KUKKONEN1979,TAKADA1993,DORNHEIM2022b}, and the exchange-correlation kernel of time-dependent DFT~\cite{Runge1984,Petersilka1996,Ullrich2012}. Indeed, static kernels constructed from QMC data have proven remarkably effective in describing dynamical and spectral properties in the electron gas~\cite{PANHOLZER2018,KOSKELO2025} and in real materials~\cite{PANHOLZER2018,CAZZANIGA2011,Chulkov2006,Pitarke2003,Patrick2015,Olsen2019}, making accurate benchmark data for $G(q)$ especially valuable.

Our knowledge of $G(q)$ is still limited, especially in the low-density regime where correlations are strong. Specifically, its behavior around intermediate wave vectors ($q/k_F\sim1$--$4$, with $k_F$ the Fermi wave vector) has remained an open question despite decades of scrutiny~\cite{GIULIANI2005,SHIRRON1986,WANG1984, SIMION2008}\footnote[2]{An indication of a local minimum appears in Ref.~\cite{SENATORE1999} based on the DMC data discussed in the main text, but the observation was limited to a single density and wave vector, and was not studied systematically.}. Yet this is the region that often matters the most: it corresponds to distances of the order of the interparticle spacing, it contains $2k_F$, where backscattering gives rise to Friedel oscillations~\cite{SIMION2005}, and it is the important region in which $G(q)$ is not fixed by exact limits.

After decades of discussion and approximate many-body schemes~\cite{HUBBARD1958,Geldart1966,Singwi1968,Geldart1970,Toigo1970,Toigo1971,Vashishta1972,KUGLER1975,Rajagopal1972,Tripathy1977,Holas1979,Devreese1980,Brosens1980,Utsumi1980,Utsumi1980b,Utsumi1981,Ichimaru1982,ORTIZ1994,Chihara1979,Mahan1981,GIULIANI2005}, the first accurate QMC data for $\chi(q)$ and $G(q)$ at zero temperature were provided by Sugiyama \textit{et al.}~\cite{SUGIYAMA1992,BOWEN1994} and Moroni \textit{et al.}~\cite{MORONI1992,MORONI1995}. These spanned the metallic range, with the dimensionless Wigner--Seitz radius $r_s = 2$--$10$, and showed that $G(q)$ varies smoothly between its asymptotic limits with a crossover near $2k_F$. They were subsequently parametrized by Corradini \textit{et al.}~\cite{CORRADINI1998} in a form that has been widely adopted in many-body calculations~\cite{OLEVANO1999,PALUMMO1999,CAZZANIGA2011,PANHOLZER2018,RAMAKRISHNA2021}, and later refined in Ref.~\cite{KAPLAN2023} using additional diagrammatic QMC data at $r_s=1,2$~\cite{KUKKONEN2021}. Dornheim \textit{et al.} performed calculations of $\chi(q)$ and $G(q)$ via path integral Monte Carlo (PIMC), but these were limited to high temperatures~\cite{DORNHEIM2020,DORNHEIM2019}.

\begin{figure*}[ht]
    \centering
    \includegraphics[width=\linewidth]{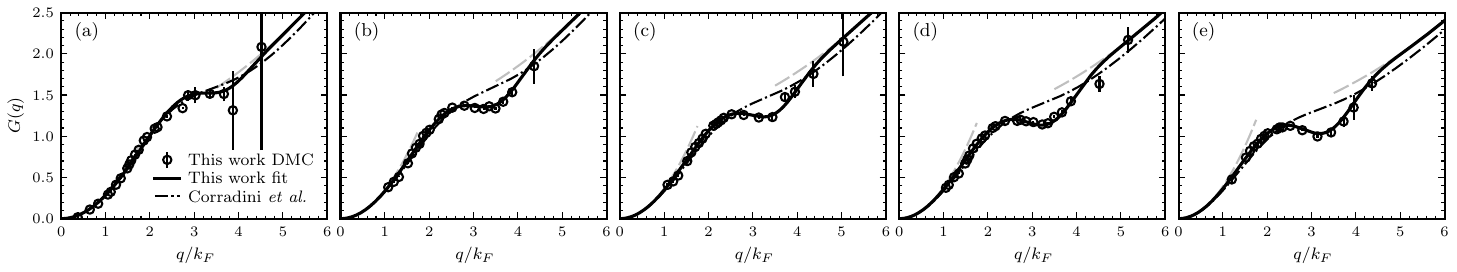}
    \caption{
        Local field factor $G(q)$ of the electron gas for (a)--(e) $r_s = 5$, $30$, $50$, $70$, and $100$. DMC data (circles) are obtained with $N_e = 162$ at $r_s = 5$ and $70$, $N_e = 54$ at $r_s = 100$, and both sizes at $r_s = 30$ and $50$. The solid line is our fit defined in Eq.~\eqref{eqn:G_fit}, and the dash-dotted line is the parametrization of Ref.~\cite{CORRADINI1998}, extrapolated. Dashed lines represent the low- and high-$q$ limits.
    }
    \label{fig:G_chi_rs20_30_50}
\end{figure*}

This leaves a critical gap: the zero-temperature QMC data for $G(q)$ of the unpolarized gas stop at $r_s = 10$. As $r_s$ increases beyond the metallic range, exchange-correlation (xc) effects become nonperturbative, and the static response function itself encodes signatures of the growing correlations, well before the eventual Wigner crystallization transition located between $r_s \simeq 87$ and $r_s \simeq 113$~\cite{AZADI2022,CEPERLEY1980, DRUMMOND2004,HOLZMANN2020}. Accurate QMC data in the strongly coupled regime $r_s>10$ are therefore highly desirable to address open questions regarding the structure of the ground-state $G(q)$, especially at intermediate wave vectors, and its physical significance to both the charge order of the electron gas and its spectral signatures in the strongly correlated liquid regime.

In this Letter, we fill this gap with systematic ground-state DMC calculations of $\chi(q)$ for the three-dimensional electron gas in the strongly coupled regime. We rely on a number of technical advances, which improve systematic accuracy, to extract the response functions reliably in this regime. Our findings show that $G(q)$ departs from a monotonic interpolation beyond the Fermi liquid regime, well before crystallization, and develops structure that reflects short-range order and reshapes the excitation spectrum. This structure is required to  reproduce the correct behavior of the structure factor $S(q)$, and it carries over to all quantities constructed from $G(q)$. By short-range order we mean the local structure of the pair correlation function $g(r)$, specifically the emergence of a peak at the edge of the xc hole around each electron, near the first-neighbor distance~\cite{DAVOUDI2003,GATTOBIGIO2005}. In the structure factor $S(q)$ (its Fourier counterpart), this appears as the main peak and the dip and oscillations that follow.

Our approach builds on the strategy of Moroni \textit{et al.}~\cite{MORONI1995}, which we extend with several technical improvements. As in their work, the system is subjected to a static perturbation $v_{\mathrm{ext}}(\mathbf r)=2v_\mathbf{q}\cos(\mathbf q\!\cdot\!\mathbf r)$, and $\chi(q)$ is computed from the leading-order response of the energy per particle, $E(v_\mathbf{q})=E_0+\chi(q)v^2_\mathbf{q}/n_0$, with $E_0$ being the unperturbed ground-state energy and $n_0=k_F^3/(3\pi^2)$ being the average density. %
The local field factor $G(q)$ is then obtained from $\chi(q)$ via the Dyson equation
\begin{equation}\label{eq:chi}
    G(q) = \frac{1}{v_c(q)} \left(\chi^{-1}(q) - \chi_0^{-1}(q)\right) + 1,
\end{equation}
where $v_c(q)$ is the Coulomb interaction and $\chi_0(q)$ is the static response function of the noninteracting system~\cite{GIULIANI2005}. The high- and low-$q$ limits of the local field factor are known~\cite{MORONI1995,CORRADINI1998,HOLAS1987,Toulouse2005}.

To compute the perturbed energies $E(v_\mathbf{q})$, we employ fixed-node DMC~\cite{CEPERLEY1978,CEPERLEY1980,MORONI1995} with Slater--Jastrow trial wave functions~\cite{JASTROW1955,LEE1981,SCHMIDT1981} implemented in \textsc{QMCPACK}~\cite{QMCPACK_1,QMCPACK_2}. Single-particle orbitals in the Slater determinant are computed using \textsc{Quantum ESPRESSO}~\cite{q-e_short}. Precise numerical determination of $\chi(q)$ is extremely computationally demanding. A straightforward workflow would involve thousands of QMC calculations at each density to determine: the maximum $v_{\bs{q}}$ of the linear-response regime, then at each $\bs{q}$ and $v_{\bs{q}}$, the optimal Slater determinant and Jastrow factor, and the finite-size error of the total energy by size extrapolation. We alleviate this cost through a three-pronged strategy. First, we exploit the noninteracting system, which both determines the linear-response window and, through its finite-size $\chi_0(q)$, supplies the size correction in Eq.~\eqref{eq:chi}~\cite{MORONI1995,GUNESprb}. Second, the orbitals are computed in a scaled potential $v^{\mathrm{eff}}_{\mathrm{ext}}(\mathbf{r})=\alpha(q,r_s)\,v_{\mathrm{ext}}(\mathbf{r})$, with the prefactor $\alpha$ optimized in variational Monte Carlo and parametrized over the whole density range of interest~\cite{GUNESprb}. This significantly reduces the computational cost of the production calculations. Lastly, we verify that using more sophisticated trial wave functions, such as backflow correlations~\cite{KWON1993,KWON1998}, on top of the optimized $\alpha$ does not affect the extracted $\chi(q)$, which depends on the curvature of $E(v_\mathbf{q})$ rather than on its value~\cite{GUNESprb}. Ground-state response calculations at strong coupling are therefore affordable with standard Slater--Jastrow wave functions and modest system sizes. Full details are given in the companion paper~\cite{GUNESprb}.

We present our results for the local field factor $G(q)$ in Fig.~\ref{fig:G_chi_rs20_30_50}. In contrast to the extrapolation of existing studies to $r_s>10$ (dash-dotted lines), where $G(q)$ is found to be smoothly evolving between its known asymptotic limits (the gray dashed lines), our findings reveal a clear structure at intermediate wave vectors. Specifically, the initially monotonic $G(q)$ is progressively depressed at intermediate wave vectors, giving rise to a suppression around $q\sim3.5k_F$. This feature becomes more prominent with increasing $r_s$, as can be seen in panels (b)--(e) of Fig.~\ref{fig:G_chi_rs20_30_50}. One would naturally expect the electron liquid to develop nontrivial behavior as it approaches Wigner crystallization, but it is more surprising that the feature appears far from it. Figure~\ref{fig:G_chi_rs20_30_50}(a) shows that it is already discernible at $r_s=5$, within the density range of real metals, and grows continuously from there.

Indications of structure at intermediate wave vectors were also seen at finite temperature~\cite{DORNHEIM2020,DORNHEIM2024}; however, with thermal and correlation effects inextricably entangled at finite $T$, a clear origin of this feature could not be established. Our DMC results agree quantitatively with these PIMC data at $\Theta = T/T_F = 0.5$, despite it being a substantial fraction of the Fermi temperature~\cite{GUNESprb}. The two are therefore the same feature: a ground-state property of the strongly coupled electron gas that is essentially unchanged up to $\Theta=0.5$.

\begin{figure}[t]
    \centering
    \includegraphics[width=\columnwidth]{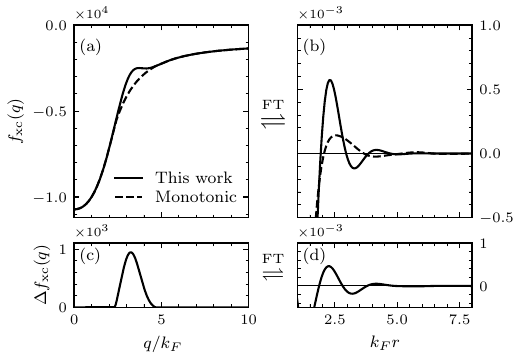}
    \caption{
        Exchange-correlation kernel $f_{\mathrm{xc}}$ at $r_s=90$. The solid line is obtained from our fitted $G(q)$, while the dashed line is obtained by manually removing the feature from the solid curve. Panel (a) shows $f_{\mathrm{xc}}$ as a function of wave vector, while panel (c) shows the difference between the solid and dashed lines in (a). Panels (b) and (d) are the Fourier transforms of (a) and (c), respectively.}
    \label{fig:fxc_r}
\end{figure}
As we will see, the structure in $G(q)$ is rooted in short-range order, and has a simple counterpart in real space. For example, the real-space exchange-correlation kernel $f_{\mathrm{xc}}(r)$, obtained by Fourier transforming $f_{\mathrm{xc}}(q) = -v_c(q)\,G(q)$, has a sharp peak near $k_Fr=2.2$ and a dip near $k_Fr=3.1$~\cite{Toulouse2005}. This strongly oscillatory feature is absent when the bump around $3k_F$ is removed from $f_{\mathrm{xc}}(q)$ (see dashed lines in Fig.~\ref{fig:fxc_r}). In other words, the structure in $G(q)$ corresponds to well-defined length scales in real space, where the density modulations cost energy around $k_Fr=2.2$ and have the opposite effect near $k_Fr=3.1$.

\begin{figure}[b]
    \centering
    \includegraphics[width=\columnwidth]{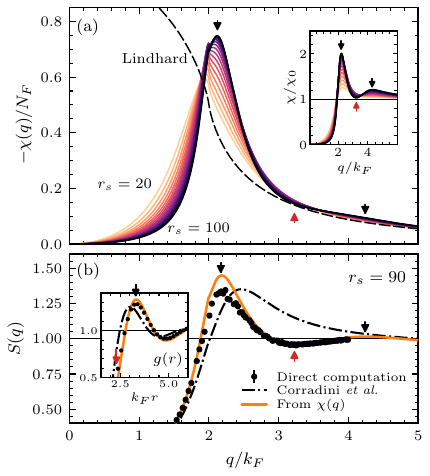}
    \caption{(a) Static density response function $\chi(q)$, in units of $N_F=k_F/\pi^2$, for $r_s=20$--$100$ from our fit using Eq.~\eqref{eqn:G_fit}; dashed: noninteracting response $\chi_0(q)$. Inset: $\chi(q)/\chi_0(q)$. Black arrows: first and second peaks of $\chi(q)$, red arrows: midpoint between the peaks. (b) $S(q)$ at $r_s=90$: DMC ($N_e=162$, circles) and SMA, scaled in order to highlight the qualitative agreement, from our $\chi(q)$ (solid) and from Corradini \textit{et al.} (dash-dotted). Black arrows: first and second peaks of $S(q)$. Inset: corresponding $g(r)$, with arrows marking the real-space counterparts (see text).}
    \label{fig:chi_sk}
\end{figure}

The observation of a well-defined xc length scale at high $r_s$ raises the question of how it relates to the incipient structural order of the electron liquid approaching Wigner crystallization. For this discussion and for broader use, it is useful to have $G(q)$ over the full range of densities and wave vectors of interest. We therefore fit our data with
\begin{equation}\label{eqn:G_fit}
    G(q) = (1-s)\,G_\mathrm{lo}(q) + s\,G_\mathrm{hi}(q), \quad s=\tanh(r_s^4/250),
\end{equation}
where both terms take the Kaplan--Kukkonen form~\cite{KAPLAN2023}
\begin{equation}\label{eqn:G_lo}
    G_\mathrm{lo,hi}(q) = \underbrace{A x^{2} h}_{\text{small }x}
    + \underbrace{\theta x^{n+2} h}_{\text{interm. }x}
    + \underbrace{(Cx^{2}+B)(1-h)}_{\text{large }x},
\end{equation}
with $x=q/k_F$, the small- and large-$q$ coefficients $A$, $B$, $C$, and $h=(e^{a_3a_4}-1)/(e^{a_3x^4/16}+e^{a_3a_4}-2)$. $G_\mathrm{lo}$ is the original form, valid for $r_s\lesssim2$: $n=4$, $\theta=a_0+a_1e^{-a_2r_s}$, with the $a_i$ of Ref.~\cite{KAPLAN2023} except $a_0=-0.00463$ and $a_2=0.623$. $G_\mathrm{hi}$ uses primed parameters, $\theta=a'_0+a'_1e^{-a'_2r_s}$ and $n=a'_5+a'_6e^{-a'_7r_s}$, with $a'_3=0.2238$, $a'_4=3+4e^{-r_s/100}$, and $a'_{0,1,2,5,6,7}=(-0.04112,\,0.03878,\,0.01607,\,2.061,\,0.903,\,0.0490)$.

Figure~\ref{fig:chi_sk}(a) shows $\chi(q)$ between $r_s=20$ and $100$, obtained from our fit of $G(q)$ to the DMC data [Eq.~\eqref{eqn:G_fit}]. As $r_s$ increases, $\chi(q)$ is suppressed toward the noninteracting response $\chi_0(q)$ at intermediate wave vectors (red arrows) before approaching the high-$q$ limit. This is the same structure as in $G(q)$, or equivalently the peak in $f_{\mathrm{xc}}(r)$, entering through Eq.~\eqref{eq:chi}. This suppression is then followed by a second peak in $\chi(q)/\chi_0(q)$, at about twice the wave vector of the first peak, more easily seen in the inset. Figure~\ref{fig:chi_sk}(b) provides a complementary view from the perspective of the structure factor at $r_s=90$. We obtain it from our fit by employing the single-mode approximation (SMA)~\cite{Bijl1940,Feynman1954}, $S(q,\omega) \simeq S(q)\delta(\omega - \omega_q)$, which, with the $f$-sum rule and the Kramers--Kronig relation at $\omega=0$, yields $S(q) = \frac{q}{2}\sqrt{-\chi(q)/n_0}$. The main peak and the following dip are located at the same wave vectors as in the inset of Fig.~\ref{fig:chi_sk}(a). Within this context, the dip reflects the electron liquid becoming stiffer against density modulations, as given by  $-1/\chi(q)$~\cite{GIULIANI2005}. Our fitted $\chi(q)$ reproduces both features (solid line), whereas a response function lacking the structure in $\chi(q)$, such as the Corradini \textit{et al.} parametrization~\cite{CORRADINI1998} (dash-dotted), cannot. This is not specific to the SMA: solving the full Dyson equation for $\chi(q)$ within the adiabatic approximation\footnote[1]{The adiabatic approximation is given by $\chi(q,\omega) = \chi_0(q,\omega) + \chi_0(q,\omega)\,v_c(q)[1-G(q)]\,\chi(q,\omega)$, from which $S(q)$ is obtained by integrating $S(q,\omega) = -\operatorname{Im}\chi(q,\omega)/(\pi n_0)$ over frequency.} yields the same correspondence~\cite{GUNESprb}. The shape of $S(q)$ is further validated by a direct computation from the equal-time pair correlation function in DMC (circles). The stiffening of $\chi(q)$ at intermediate wave vectors is required to reproduce both the location of the main peak of $S(q)$ and the dip that follows. The fact that this is short-range order seen through the static response is highlighted by the pair correlation function $g(r)$ in the inset of Fig.~\ref{fig:chi_sk}(b). This inset shows $g(r)$ at $r_s=90$ in real space, comparing the result of a direct computation to the $g(r)$ obtained from the Corradini extrapolation and from our fit to the new data. The comparison shows that only the most recent data and fit give the correct peak position (black arrow) and steepness of the rise starting from small distances. This pronounced peak with its steep rise at the edge of the exchange-correlation hole reflects the tendency of the system to establish short-range order, and our analysis demonstrates in which way the response properties, and in particular $G(q)$, contribute to this phenomenon. One may also note the relation to $f_{\mathrm{xc}}(r)$ in Fig.~\ref{fig:fxc_r}, with the position of its maximum given by the red arrow in the inset of Fig.~\ref{fig:chi_sk}(b).

\begin{figure}[t]
    \centering
    \includegraphics[width=\linewidth]{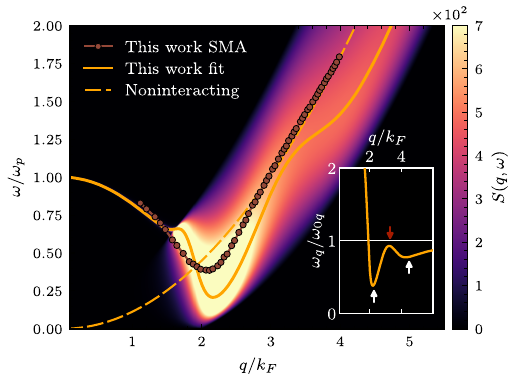}
    \caption{Dynamic structure factor $S(q,\omega)$ for $r_s=90$ calculated using the static $G(q)$ as the xc kernel in the linear-response $\chi(q,\omega)$. Solid orange line: collective mode dispersion from the peak of $S(q,\omega)$. Dashed line: the noninteracting dispersion, $\omega_{0q} = q^2/2$. Data points: collective mode dispersion using DMC $S(q)$ as input for SMA, i.e., $\omega_q = q^2 / [2S(q)]$. Inset: the solid orange line in the main panel divided by $\omega_{0q}$. The white and red arrows mark the dips and peaks, respectively.}
    \label{fig:dynamic_sk}
\end{figure}

Another important application of our parametrization is to access dynamical properties. To this aim, we compute the dynamic structure factor $S(q,\omega)$ at $r_s=90$ within the adiabatic approximation, i.e., replacing the dynamic local field factor in $\chi(q,\omega)$ by the static $G(q)$~\footnotemark[1]. The collective mode, identified with the peak of $S(q,\omega)$ in Fig.~\ref{fig:dynamic_sk}, is a well-defined plasmon at small $q$, and acquires a finite Landau width as it enters the particle-hole continuum. In contrast to the Fermi liquid regime where the plasmon dissolves in the continuum, we find a low-frequency resonant mode that persists and develops a minimum around $q \sim 2.2k_F$. The same minimum follows from the SMA applied to our DMC $S(q)$ [Fig.~\ref{fig:chi_sk}(b)], an equal-time quantity obtained from a separate DMC calculation. Therefore, this mode is already encoded in the independently computed structure factor. The effect of the structure in $G(q)$ is further illustrated in the inset, which shows the collective mode dispersion normalized by $\omega_{0q}=q^2/2$. The first dip lies at $q\approx2.2k_F$, a second at twice that wave vector, $q\approx4.4k_F$, with a local maximum in between. The dips mirror the peaks of $\chi(q)$ and $S(q)$ in Figs.~\ref{fig:chi_sk}(a) and~\ref{fig:chi_sk}(b) (black arrows), while the maximum reflects the structure in $G(q)$ (red arrows), which also produces the second dip. The static correlations therefore survive in the dynamic response, which carries direct fingerprints of both the peak at $2.2k_F$ in $\chi(q)$ and the dip around $3.2k_F$ in $G(q)$.

The underlying physics of this low-energy mode has been investigated: a resonant excitonic mode inside the particle-hole continuum was proposed by Takada within a dielectric formalism~\cite{TAKADA2016}, and recently obtained from the Bethe--Salpeter equation~\cite{KOSKELO2025}. Our results are consistent with these studies. The mode follows from $G(q)$ exceeding unity for $q \gtrsim 2k_F$ and causing the net particle-hole interaction to become attractive. Both earlier approaches relied on extrapolated or approximate local field factors due to a lack of QMC data; our $G(q)$ supplies that input and enables revisiting these studies throughout the strongly coupled regime. Furthermore, a collective mode that remains visible throughout the whole continuum also has an experimental precedent in a two-dimensional Fermi liquid, where it is characterized by a roton-like minimum~\cite{GODFRIN2012}, while the second minimum of $\omega_q/\omega_{0q}$ has recently been interpreted as a second roton feature in finite-temperature PIMC simulations at large $r_s$~\cite{CHUNA2025}. Our findings, together with the PIMC data, thus show this to be an intrinsic zero-temperature phenomenon that persists at finite temperature. The name ``roton'' is inherited from superfluid helium; it characterizes the shape of the dispersion, but not its microscopic origin. Though we only show $r_s=90$ in Fig.~\ref{fig:dynamic_sk}, the depth of the minimum grows with $r_s$ as the liquid is driven toward the crystallization regime, which is consistent with the picture in which such roton-like features signal the incipient order of a strongly coupled liquid approaching Wigner crystallization~\cite{GODFRIN2012,NOZIERES2004,PARK2024}. Beyond the illustration above, our parametrization provides the input needed for vertex corrections beyond $GW$~\cite{DELSOLE1994,VERGNIORY2007,SCHMIDT2017} and for extending the response calculations of warm dense matter to lower temperatures~\cite{DORNHEIM2019}.

In summary, we have presented benchmark DMC calculations of the static density response function $\chi(q)$ of the three-dimensional electron liquid, from metallic densities to the vicinity of Wigner crystallization. Extracted from $\chi(q)$, the local field factor $G(q)$ shows a pronounced structure that signals the buildup of short-range order, simultaneously reflected in the structure factor and exchange-correlation kernel. Our results thus settle the long-standing question of the electron gas response at intermediate wave vectors and yield a complete parametrization of $G(q)$ over the full density range of the liquid. At metallic densities it supersedes the forms currently in use, which do not capture this structure; at low densities it yields a resonant collective mode that survives inside the particle-hole continuum with a roton-like minimum, consistent with the low-energy excitonic mode found in recent many-body spectroscopy calculations. Intermediate wave vectors are where correlation leaves its clearest signature. Our parametrization makes this regime available to the construction of nonlocal exchange-correlation functionals, in which $G(q)$ enters directly, and motivates further research in other quantities, such as the spin response, and in two dimensions~\cite{SMITH2024}, where the liquid crystallizes at far higher density and the interplay of correlations and dimensionality may lead to new phenomena.

\begin{acknowledgments}
    The authors would like to thank Saverio Moroni, Carl Kukkonen, and Carlos Rodriguez Perez for helpful discussions. The Flatiron Institute is a division of the Simons Foundation. This work has received state funding managed by the French National Research Agency (ANR) under the France 2030 program, reference ``Grant No.~ANR-22-EXES-0013''. YY acknowledges support by the National Science Foundation (NSF) under grant number DMR-2532734. This work used Delta through allocation PHY250273 from the Advanced Cyberinfrastructure Coordination Ecosystem: Services \& Support (ACCESS) program, which is supported by U.S. National Science Foundation grants \#2138259, \#2138286, \#2138307, \#2137603, and \#2138296.
    This research used the Delta advanced computing and data resource which is supported by the National Science Foundation (award OAC 2005572) and the State of Illinois. Delta is a joint effort of the University of Illinois Urbana-Champaign and its National Center for Supercomputing Applications.
\end{acknowledgments}

The data that support the findings of this article are openly available~\cite{zenodo}.

\bibliographystyle{apsrev4-2}
\bibliography{main}

\end{document}